# Out-of-time-order correlators bridge classical transport and quantum dynamics

Sophia N. Fricke,[1*] Haiyan Mao,[2] Manas Sajjan,[3] Ashok Ajoy,[4,5,6] Velencia Witherspoon,[7] Sabre Kais,[3,8] Jeffrey A. Reimer[1,9]

[1]Pines Magnetic Resonance Center, University of California Berkeley, Berkeley, CA 94720, USA

[2]Department of Materials Science and Engineering, Stanford University, Stanford, CA 94305, USA

[3]Department of Electrical and Computer Engineering, North Carolina State University, Raleigh, NC 27606, USA

[4]Department of Chemistry, University of California Berkeley, Berkeley, CA 94720, USA

[5]Chemical Sciences Division, Lawrence Berkeley National Laboratory, Berkeley, CA 94720, USA

[6]CIFAR Azrieli Global Scholars Program, 661 University Ave, Toronto, ON, M5G 1M1, Canada

[7]Department of Biomedical Engineering, Tulane University, New Orleans, LA 70118, USA

[8]Miller Institute for Basic Research in Science, 468 Donner Lab, Berkeley, CA 94720, USA

[9]Department of Chemical and Biomolecular Engineering, University of California Berkeley, Berkeley, CA 94720, USA

[*]Corresponding Author, **Email:** snfricke@berkeley.edu

**Author contributions:** S.N.F. designed, performed, and analyzed research and wrote the paper, H.M. synthesized and characterized materials, M.S. and S.K. helped develop the OTOC derivation, M.S. and A.A. and V.W. edited the paper, and J.A.R. designed and directed the research and edited the paper.

**ABSTRACT.** The out-of-time-order correlator (OTOC) has emerged as a central tool for quantifying decoherence across wide-ranging physical platforms. Here we demonstrate its direct measurement in a classical ensemble using nuclear magnetic resonance (NMR) with a modulated gradient spin echo (MGSE) sequence and extend the method into a multidimensional correlation to track exchange phenomena. Position is encoded through magnetic field gradients and momentum through the velocity autocorrelation function, enabling experimental access to OTOCs for proton motion confined within the self-similar lattice of the metal–organic framework MOF-808. Here, water confined to specified geometries within the MOF pores gives rise to spatially distinct diffusive eigenmodes with characteristic relative entropies. We demonstrate that periodic radiofrequency (rf) driving combined with gradient modulation yields entropy evolution through the selection of distinct diffusion modes. Frequency-resolved diffusion spectra connect these entropy dynamics to classical heat exchange laws, revealing how operational features of quantum systems are mirrored in confined, macroscopic spin ensembles.

**Keywords**: out of time order correlator, time-dependent diffusion, velocity autocorrelation function, nuclear magnetic resonance, quantum dynamics

## I. INTRODUCTION

Quantum technologies are typically built on experimentally fragile, microscopic platforms—trapped ions, NV centers, or superconducting circuits—that permit exquisite control but require cryogenics, vacuum systems, and complex infrastructure [1]. Alternatively, there is increasing interest in studies of classical systems whose dynamics admit similar mathematical descriptions as quantum systems [2]. By leveraging hierarchical molecular architectures, it is possible to design macroscopic systems that offer tunability, reproducibility through chemical synthesis, and compatibility with existing technologies. Such systems create accessible experimental platforms to test theory and potentially reveal universal physics.

Metal–organic frameworks (MOFs) provide especially promising platforms for this purpose [3,4]. Their recursively self-similar crystalline architectures can confine molecules across length scales from internal pores (~1 nm) to the crystallite (>100 nm), creating hyperuniform environments where physical and spin diffusion admit mathematically equivalent descriptions [5–10]. Protons can be localized in chemically and topologically distinct settings defined by the MOF fractal crystalline structure, making MOFs versatile platforms for studying dynamics governed by conservation laws. The scale-invariant symmetry of these architectures invites comparison to Noether's theorem, which links spatial symmetries to conserved dynamical quantities [11,12]. Such conserved quantities are essential both for encoding quantum information and for describing classical transport: classically they include energy, charge, and angular momentum, while in the present context they correspond to translation-invariant Hamiltonians defined over the MOF-relevant length scales. Under these conditions, the classical transport of confined water molecules provides access to the dissipation of conserved quantities in the underlying quantum spin system [13,14].

Here, we directly probe these dynamics using the out-of-time-order correlator (OTOC), a measure of information scrambling and decoherence, via nuclear magnetic resonance (NMR) with a modulated gradient spin echo (MGSE) sequence [15–24]. In this framework, magnetic field gradients encode the spatial position of aqueous protons, while velocity autocorrelations encode momentum, enabling room-temperature measurement of the OTOC in water confined within the zirconium-based MOF-808 [25,26]. A multidimensional extension enables the study of water exchange dynamics between reservoirs whose fixed geometries impose characteristic diffusive eigenmodes. Periodic radiofrequency (rf) driving, interspersed with variable delays, selects distinct diffusion eigenmodes and allows entropy changes to be tracked over timescales extending to days [27]. By monitoring exchange correlations and incremental entropy changes through frequency-resolved diffusion spectra calculated from OTOCs, we demonstrate that confined, macroscopic water can reproduce operational features typically associated with quantum systems.

## II. METHODS

### A. Materials

The MOF-808(Zr) compound, selected for its thermal and chemical stability, was synthesized in a manner similar to that previously reported. [28,29] Here, 0.315 g of trimesic acid (H3BTC - 95 %, Sigma Aldrich) and 0.727g of zirconium chloride (ZrOCl2.8H2O - 98 %-Sigma-Aldrich) were added to 33.7 mL of DMF (HCON(CH3)2 – 99.8 %, Sigma Aldrich) and 33.7 mL of formic acid (HCOOH - 95 %, Sigma Aldrich), and the solution was subjected to 10 minutes of sonication. After this step, the mixture transferred to a Teflon-lined autoclave, and heated to 130 °C, allowing the reaction to proceed for a duration of 48 hours. Subsequently, the solution was cooled to ambient temperature, and the solution was filtered and washed three times with DMF (30 mL), deionized water (30 mL) and acetone (30 mL) (CH3COCH3 – 99.5 %, Sigma Aldrich) each. The solid was dried at 150 °C for 24 hours. A powder x-ray diffraction pattern is provided in the Supplemental Material (Fig. S1). Celery was used as a well-established and readily accessible control sample for one-dimensional diffusion [30–33].

### B. Experimental Design

NMR modulated gradient spin echo (MGSE) [22,23] measurements were performed at room temperature with an NMR-MOUSE (Mobile Universal Surface Explorer) PM25 0.3 T unilateral magnet [34,35] and a Magritek Kea II spectrometer at a $^{1}$H resonant frequency of 13.11 MHz with a constant gradient of 7 T m$^{-1}$. The number of π rf pulses in a Carr–Purcell–Meiboom–Gill pulse sequence (CPMG) [36,37] was varied within a fixed total time of 55 ms per echo train to selectively detect the diffusive contribution to signal relaxation, and not transverse relaxation, using Prospa v3.61 software from Magritek (Malvern, PA.). For all experiments, π/2 rf pulse lengths were 2.5 μs, the delay between π rf pulses was varied between 55 and 1100 μs in 20 incremental steps for both the direct and the indirect dimension, typical mixing times varied between 0.5–10 ms, and the repetition time for signal averaging was 2.4 s to sum 128 CPMG transient signals. To cancel artifacts arising from pulse imperfections, the initial π/2 rf pulse and the receiver were cycled between +x and −x phase, while holding the π rf pulse phase constant at +y. Calculation of $D(\omega)$ from MGSE experiments was performed using a formula for echo attenuation (corresponding to the OTOC, $F(t)$) as a function of gradient modulation frequency, $E(\tau, \omega_m) = \sum_i E_{0,i} e^{-\frac{\tau}{T_2} - \frac{8\gamma^2 G^2}{\pi^2 \omega_m^2} D_{zz,i}(\omega_m)\tau}$ [22,23].

For the entropy modulation experiment, echo times were incremented periodically from 55.00, 68.75, 91.67, 137.50, 91.67, 68.75, 55.00 μs, with the number of echoes varied from 1000, 800, 600, 400,

600, 800, 1000 respectively; this sequence was repeated for twenty cycles with a repetition time of 9.5 s. All data analysis was accomplished with MatLab (Mathworks, Natick, MA).

## III. RESULTS

### A. OTOC for spin diffusion

The out of time order correlator (OTOC) is a theoretical and experimental framework for quantifying decoherence and information scrambling in systems ranging from quantum circuits to black holes. It is formally defined as

$$C(t) \equiv -\langle [W(t), V(0)]^2 \rangle \tag{1}$$

where $W(t)$ and $V(t)$ are any Hermitian pair of local, time-dependent Heisenberg operators averaged over a given system [15–18]. In the present context, these operators are elements of the SU(2) Lie algebra generated by spin operators, which exhibit a continuous U(1) subgroup corresponding to phase rotations about the quantization axis. This U(1) symmetry underlies conservation of total spin projection in the absence of relaxation, linking the OTOC to experimentally accessible quantities such as the velocity autocorrelation function in a magnetic field gradient.

As a dynamic observable, the OTOC quantifies the growth of noncommutativity between initially local operators, reflecting how localized information becomes delocalized—or "scrambled"—across the Hilbert space of a quantum many-body system. This operator spreading of initially localized information is central to diagnosing quantum chaos, as it captures the rate and manner in which perturbations propagate through the system. Originally introduced in the context of higher-order corrections to current response in superconductors [38], the OTOC has since found wide-ranging applications: detecting many-body localization and dynamical phase transitions with non-analyticities in Loschmidt echoes [39,40], analyzing information flow in quantum machine learning models [41], and bounding the scrambling capacity of black holes via anti-de Sitter/conformal field theory (AdS/CFT) duality, which has advanced the current understanding of string theory and quantum gravity [42–44]. Recent algorithmic developments have enabled OTOC evaluation on quantum circuits via interferometric protocols [45], randomized measurements and classical shadows [46], making it feasible to study scrambling in highly entangled quantum states.

When $W(0)$ and $V(0)$ are unitary and Hermitian, it can be shown that Eq. (1) reduces to $C(t) = 2(1 - Re\langle W(t)V(0)W(t)V(0)\rangle)$. In these cases, the quantity $\langle W(t)V(0)W(t)V(0)\rangle$ is of central importance to characterize OTOCs and is often analyzed exclusively [47]. Here, we focus on a specific operator choice: $W(0) = \hat{x}(0)$, the position operator, and $V(0) = \hat{p}(0)$, the canonically conjugate momentum, and introduce an experimental paradigm for evaluating

$$F(t) = -\langle x(t)p(0)x(t)p(0)\rangle \tag{2}$$

using nuclear magnetic resonance (NMR) (vide infra). This pair of operators is historically well-studied due to their classical analog, evaluated via the dequantization procedure of converting commutators to Poisson brackets: $i\hbar[,] \rightarrow \{,\}_{PB}$ [48,49]. Here, exponential divergence of nearby trajectories is quantified by the expression

$$\left|\frac{\delta x(t)}{\delta p(0)}\right| = \left|\{x(t), p(0)\}_{PB}^2\right| \sim e^{2\lambda_L t} \tag{3}$$

where $\lambda_L$ represents the Lyapunov exponent [16–18]. Thus, in semi-classical regimes, the OTOC defined in Eq. (1) is readily shown to approach a form given by $F(t) \sim e^{2\lambda_L t}$, from which Lyapunov exponents can be extracted [50]. This framework naturally provides an operational definition for diagnosing quantum chaos, where operator growth—not phase-space trajectories—encodes the scrambling dynamics. The spatial propagation of scrambling is characterized by the “butterfly” velocity, obtained from the spacetime profile of $F(t)$ (or more generally, of $C(t)$) as the slope of the contour defining the boundary of the commutator core. This velocity quantifies the speed at which initially localized perturbations influence distant degrees of freedom, complementing the Lyapunov exponent that describes the temporal growth rate. In generic chaotic systems with short-range interactions, sharply localized initial operators and open or weakly confining boundaries tend to produce a clear regime of short-time exponential growth in $F(t)$. In contrast, extended or delocalized initial conditions, strong confinement, or boundaries that reflect operator growth can suppress or obscure the exponential regime, making the observed behavior highly sensitive to these system-specific attributes [16–18].

**B. OTOC for physical diffusion**

It is convenient to recognize that the OTOC of position and momentum can be recast in terms of a time-dependent diffusion measurement that can be performed with NMR using a modulated gradient spin-echo (MGSE) sequence. In this framework, position is encoded by a magnetic gradient-time wave vector and momentum is encoded by the velocity autocorrelation function of spin-bearing molecules [51,52]. Here, the rotating frame Hamiltonian for non-interacting spins may be written as

$$\hat{H}(t) = -\hbar \sum_i \left(\omega_0 \hat{I}_{z,i} + \omega(\vec{r}_i)\hat{I}_i\right) + \hat{H}_{rf}(t) \tag{4}$$

where $\omega_0 = \gamma \vec{B}_{0,z}$ represents the Zeeman interaction and $\omega(\vec{r}_i) = \gamma \underline{G}\vec{r}_i$ represents the spatially dependent frequency offsets resulting from a magnetic field gradient, $\underline{G}$, $\gamma$ is the magnetogyric ratio, and the summation in the first term is over the $i$ spins in the ensemble represented by the spin operator $\hat{I}$. The second term represents the interaction of an rf pulse with the spin system, which is here given by

$$\hat{H}_{rf}(t) = \hat{H}^x_{\pi/2}(t) + \hat{H}^y_{CPMG}(t) \tag{5}$$

for an echo train experiment designed to measure the OTOC for time-dependent diffusion in a closed system (Fig. 1). The second term is

$$\hat{H}^y_{CPMG}(t) = -2\omega_\pi(t)\cos(\omega_0 t)\sum_i \hat{I}_{y,i} \tag{6}$$

where $\omega_\pi$ is the frequency of $\pi$ rf pulses with amplitude $2\omega_n/\gamma$. Transformation of Eq. (4) into the toggling frame gives

$$\hat{H}(t) = -\hbar \sum_i \omega_z(\vec{r}_i(t))\left[\hat{I}_{z,i}\cos b(t) - \hat{I}_{x,i}\sin b(t)\right] \tag{7}$$

where the effect of successive $\pi$ rf pulses is to modulate $b(t)$ from $\pm\pi$. The echo signal corresponds to the trace

$$E(t) = \hbar\gamma \frac{d}{dt}\sum_i Tr\hat{\rho}(t)\hat{I}_{y,i} \approx \sum_i \langle e^{i\int_0^t \omega_z\left(\vec{r}_i(t')\right)\cos b(t')dt'}\rangle. \tag{8}$$

When sampled stroboscopically, the peak echo amplitude at points occurring at multiples of the cycle time, $t = NT$, is

$$E(t) = \sum_i \langle e^{i\int_0^t \nabla\omega_z\left(\vec{r}_i(t')\right)\cdot\vec{v}_i(t')\cdot f_\pi(t')dt'} \rangle. \tag{9}$$

where $f_\pi(t)$ is a modulating function that alters the orientation of the magnetic field gradient in space.

The Einstein definition of diffusion considers $D$ as a time derivative of molecular mean squared displacement in the long time, or zero frequency, limit [27,53–55]. In a finite time interval, though, $D$ may often exhibit time-dependence, as the initial velocity, $\vec{v}(0)$, may remain imprinted upon motion at later times, $\vec{v}(t)$. To characterize such a quantity, we construct a rank-4 tensor, $T$, with indices $(i, j)$ indicating individual spins and $(\alpha, \beta) \in S = \{(a, b) \mid a \in \{x, y, z\}, b \in \{x, y, z\}\}$ denoting spatial components:

$$T_{\alpha,\beta}^{ij} = \int_0^\tau \left\langle v_\alpha^i(t) v_\beta^j(0) \right\rangle dt. \tag{10}$$

Here we note that the diagonal elements of this tensor yield directional time-dependent self-diffusion coefficients, such as $D_{zz}(\tau)$, corresponding to the Green–Kubo relation:

$$T_{\alpha=z,\beta=z}^{ii} = D_{zz}(\tau) = \int_0^\tau \langle v_z(t) v_z(0) \rangle_\tau dt = \frac{2}{\pi} \int_0^\infty D_{zz}(\omega)_\tau \frac{\sin(\tau\omega)}{\omega} d\omega \tag{11}$$

where movement is considered the ensemble average of spin trajectories along $\hat{z}$ within the time interval, $\tau$ [27]. The corresponding power spectrum is the velocity autocorrelation function

$$T_{zz}^{ii} = D_{zz}(\omega)_\tau = \int_0^\tau \langle v_z(t) v_z(0) \rangle_\tau \cos(\omega t)\, dt. \tag{12}$$

To describe one-dimensional diffusion spectra from an NMR measurement, the Green–Kubo relations may be associated with phase decoherence encoded by the gradient-time wave vector, $\underline{q}(t) = \gamma \int_0^t \underline{G}(t') f_\pi(t') dt'$, where $\gamma$ is the magnetogyric ratio of the detected nuclear spin and $\underline{G}(t)$ is the magnetic field gradient. In the presence of a constant magnetic field gradient, the application of $\pi$ rf pulses causes $f_\pi(t)$ to switch between $\pm 1$. This may be expressed mathematically through an oscillatory function, where $f_\pi(t) = \cos\left(b(t)\right)$ and successive $\pi$ rf pulses modulate $b(t)$ from $\pm\pi$. In the limit that molecular displacements are less than $\left|\underline{q}\right|^{-1}$ within the $\pi$ pulse-generated phase modulation interval, i.e., the diffusive regime, the Gaussian approximation may be used to express the attenuation of the echo train through a cumulant series as

$$E(\tau) = \sum_i E_{0,i} e^{-i\alpha_i(\tau) - \beta_i(\tau)} \tag{13}$$

where the nuclear spins with differing dynamics are grouped as $i$ in the sum and $\tau$ is half the time between successive $\pi$ rf pulses. Critically, the imaginary part of the phase shift,

$$\alpha_i(\tau) = \int_0^\tau \underline{q}(t) \cdot \langle \vec{v}_i(t) \rangle dt \tag{14}$$

may be neglected when the velocity of the detected molecules averages to zero [22,25]. In this case, the echo attenuation may be calculated through the real part only,

$$\beta_i(\tau) = \frac{1}{\pi} \int_0^\infty \underline{q}(\omega, \tau) \cdot \underline{D_i}(\omega, \tau) \cdot \underline{q}^*(\omega, \tau) d\omega. \tag{15}$$

Here, $\underline{q}(\omega, \tau)$ represents the wave vector that has been encoded with spin phase decoherence resulting from thermal fluctuation-driven displacements in the medium. For this case, the power spectrum of the velocity autocorrelation function from Eq. (12) is modified to

$$T_{zz}^{ii} = \underline{D_i}(\omega, \tau) = \int_0^\infty \langle \vec{v}_i(t) \otimes \vec{v}_i(0) \rangle_\tau \cos(\omega t)\, dt. \tag{16}$$

By substituting Eq. (16) into Eq. (15), it is clear that measurement of $\beta_i(\tau)$ from an MGSE experiment directly probes $F(t)$ (Eq. (2)) as the OTOC for physical diffusion in a magnetic field gradient across a many-body network.

Here, this network is realized by water molecules confined within MOF-808, a crystalline zirconium-based metal–organic framework with 0.48 and 1.84 nm pores and high thermal and chemical stability. The periodic, self-similar pore arrangement of MOF-808 provides a spatially isotropic but topologically ordered environment at length scales relevant to both spin phase evolution and physical diffusion, ensuring that transport is governed by uniform geometric constraints in all directions. These structural features satisfy the requirement for a highly ordered isotropic spatial environment, while allowing pore topology and connectivity to impose well-defined confinement effects on the spatial arrangement of spins [56]. The resulting molecular spin system, structured across multiple length scales, is depicted in Fig. 1 for aqueous protons confined in MOF-808. Such confinement naturally gives rise to diffusion eigenmodes, which provide a modal basis for describing transport under geometric constraints. These modes arise when boundary conditions quantize diffusive motion and can be calculated from the OTOC, $F(t)$.

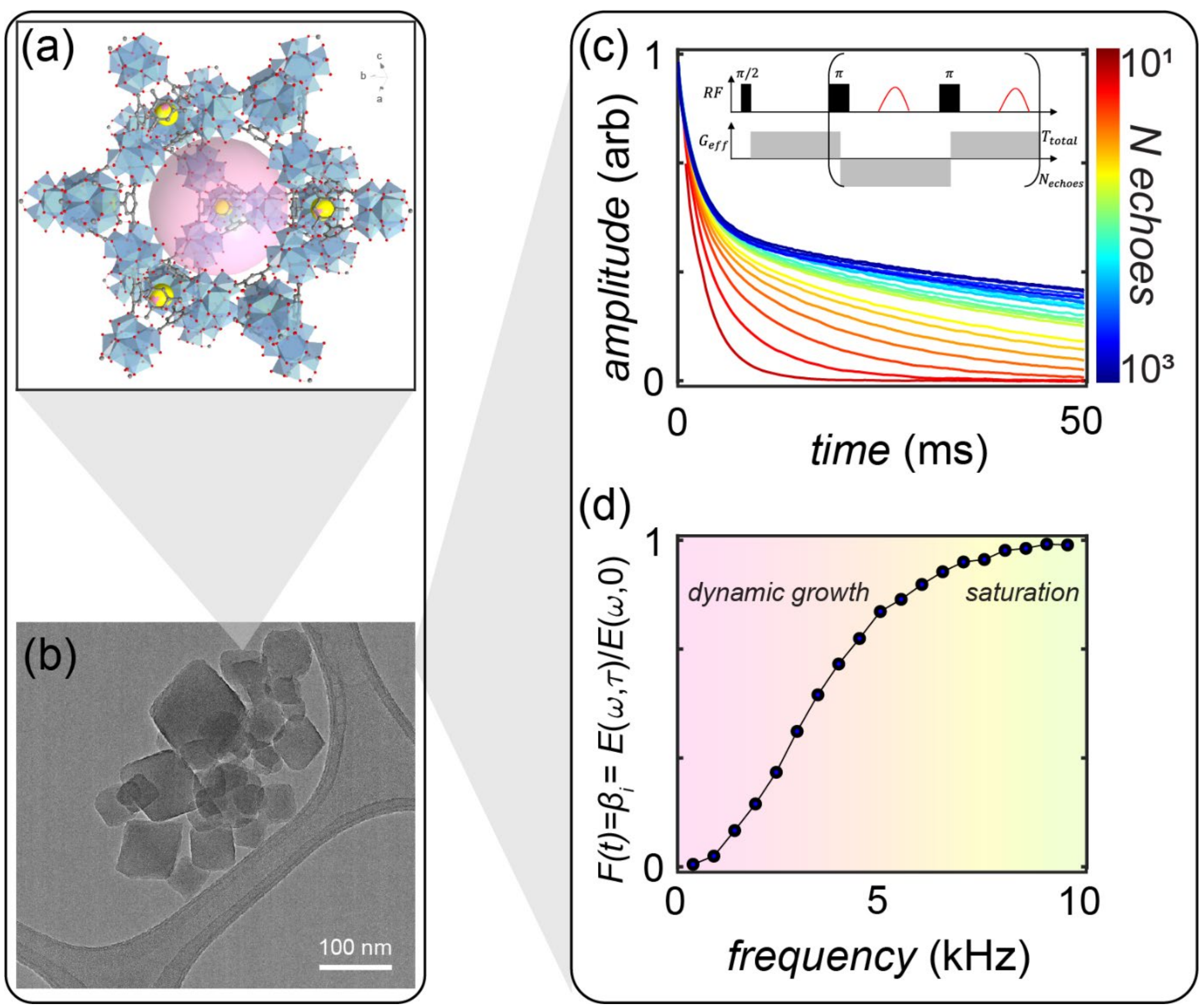

**FIG 1. Water confined in MOF-808 as a molecular spin system studied with OTOCs measured by MGSE NMR.** (a) Chemical structure of MOF-808, showing its multi-scale hexagonal framework for proton confinement with 0.48 and 1.84 nm pores. (b) TEM image highlighting the crystalline ordering and pore arrangement. (c) MGSE pulse sequence used to measure the OTOC (inset), with echo trains collected by varying the π-pulse frequency within a fixed total time to separate relaxation and diffusion effects. (d) The OTOC, $F(t)$, extracted from the normalized echo attenuation as a function of π-pulse frequency, corresponding to Eq. (15).

To probe interactions beyond frequency-resolved mode dynamics, we introduce a multidimensional extension of the OTOC (2D OTOC), which maps correlated motion across distinct timescales and length scales. This approach extends a modulated-gradient echo train to an additional dimension, using a unilateral magnet to supply a constant field gradient [25,26]. The resulting spatiotemporal correlations capture how structurally distinct domains are dynamically connected by mapping chemical structures to network dynamics, thereby linking interactions of nuclear spins to distinct chemical environments. Depending on sample preparation, these 2D OTOCs can reveal specific forms of frequency-correlated exchange across boundaries that constrain diffusion.

By monitoring frequency-correlated exchange within varying timescales, it is possible to observe the exchange of water molecules between pores, where each pore is characterized by unique diffusion dynamics. (Fig. 2). Physically, exchange between two frequencies indicates mutually diffusing molecules between differing regimes of porous confinement (localization). Analysis of multidimensional OTOCs in this way links proximally diffusing molecular moieties in dynamic settings and resolves distinct yet interfering eigenmodes of components. These insights provide assessments of the influence and extent of microstructure on the lifetimes and entanglement-like processes mediated by diffusive exchange between topologically-ordered water molecules in these systems.

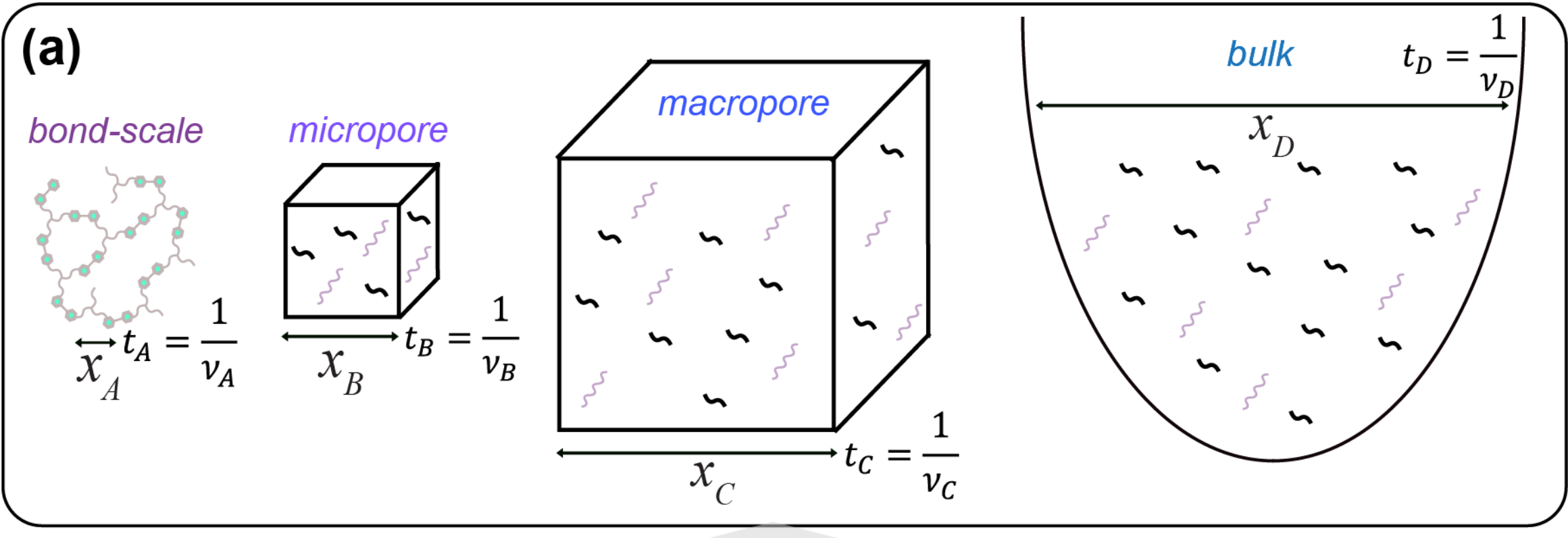

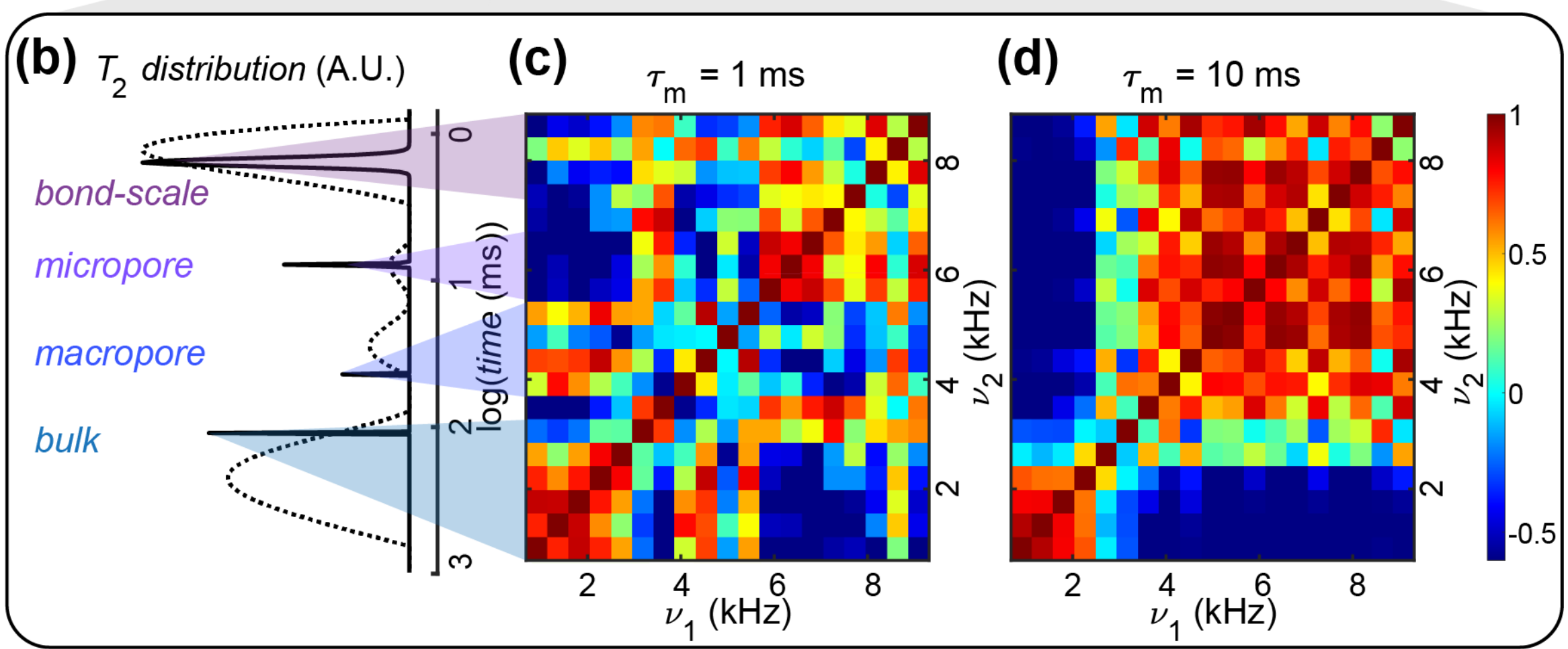

**FIG 2. Multidimensional OTOCs report correlations via spatiotemporal exchange networks in water confined in MOF-808.** (a) Schematic of water molecules experiencing motion in four distinct domains: bond-scale, micropore, macropore (intercrystallite), and bulk water. (b) $T_2 = \frac{1}{R_2}$ relaxation time distributions obtained by Laplace inversion (dashed) and matrix pencil method (solid). Peaks correspond to the four domains in (a), spanning over three orders of magnitude in relaxation times [57–60]. (c) 2D OTOC correlation plot for a 1 ms mixing time, showing discrete off-diagonal exchange between specific domains, e.g., between macropore and backbone-confined water (~2 and ~9 kHz). (d) 2D OTOC correlation plot for a 10 ms mixing time, showing widespread off-diagonal exchange, indicating extensive proton migration across domains.

Figure 2 demonstrates one implementation of this strategy to track exchange dynamics for water molecules confined at bond-scale, micropore, macropore, and bulk domains. Figure 2(b) shows distributions of relaxation times in a system with multiple domains of varying heterogeneity, as illustrated in Fig. 2(a), and how these components relate to the 2D OTOC correlation plots demonstrating exchange between these domains at varying timescales (Fig. 2(c) and 2(d)). Each of the represented domains is characterized by unique and resolved time- and length-scales. At low frequencies (< 2 kHz), the dynamics of bulk water are predominantly observed. As frequencies increase (~2–4 kHz), transient interactions become evident as scattered correlations with intercrystallite (or macro-pore) regions characterized by micron- to 100 nm-scale confinement (depicted in Fig. 1(b)). At frequencies of ~5–7 kHz, there is increasing influence from micro-pore regions with greater confinement (shown in Fig. 1(a)). Contribution from network-bound protons with the greatest confinement dominates at the highest frequencies > 8 kHz.

Measurement of exchange occurring within varying timescales of 1 ms (Fig. 2(c)) versus 10 ms (Fig. 2(d)) via 2D OTOCs demonstrates the effects of dissipation and confinement in this physical network. The more localized exchange regions in Fig. 2(c) represent "jumps" of a proton from one environment to another (e.g., between an intercrystallite region and the MOF-808 backbone), indicated by symmetric off-diagonal intensity c.a. 2- and 9 kHz, for this example. Within the 1 ms mixing time, exchange is resolved into relatively discrete off-diagonal regions, indicating that most protons have migrated between at least two environments in the crystalline network within this time (an annotated overlay is provided in Supplemental Fig. S2). Contrastingly, with a longer exchange time of 10 ms (Fig. 2(d)), more extensive mixing is observed, indicating more complete migration of protons throughout the network. Here, exchange is widespread, showing extensive movement or exchange of protons throughout the material architecture. In this framework, the observed exchange corresponds to correlations that are naturally described by a two-time OTOC, $F(t_1, t_2)$, whose frequency-domain representation is shown here and captures both the diffusion timescale and the exchange interval.

Previous work has corroborated frequency-correlated exchange with simulated data and benchmarked the technique using homopolymers with known domains arising from topology as standards [25,26]. To benchmark the multidimensional MGSE OTOC approach against a well-defined one-dimensional geometry, we performed comparative measurements on a control phantom consisting of the xylem bundles of celery, which contain parallel cylindrical water channels approximately 5 μm in diameter. By orienting the phantom either parallel or perpendicular to the applied magnetic field gradient, we can distinguish Brownian motion along the channels from confinement effects in the transverse plane. To confirm the multidimensional MGSE OTOC accurately reports physical mixing of spins initially prepared in superposition states from a π/2 pulse, Fig. 3 shows contrasting MGSE OTOC correlations of water confined in cylindrical 5 μm pores oriented (see Methods section) along the gradient dimension (Z) and orthogonal to it (XY, Fig. 3(d)) at the longest-considered mixing time of 10 ms. When oriented along Z (Fig. 3(c)), it is apparent that the primary nature of proton movement is Brownian (e.g., bulk-water like), whose signature is exchange at predominantly low (or zero) frequencies. Contrastingly, when the sample is reoriented in the magnetic field such that the cylindrical pores are in the XY plane, the effects of confinement become evident with more extensive high-frequency exchange [30–33].

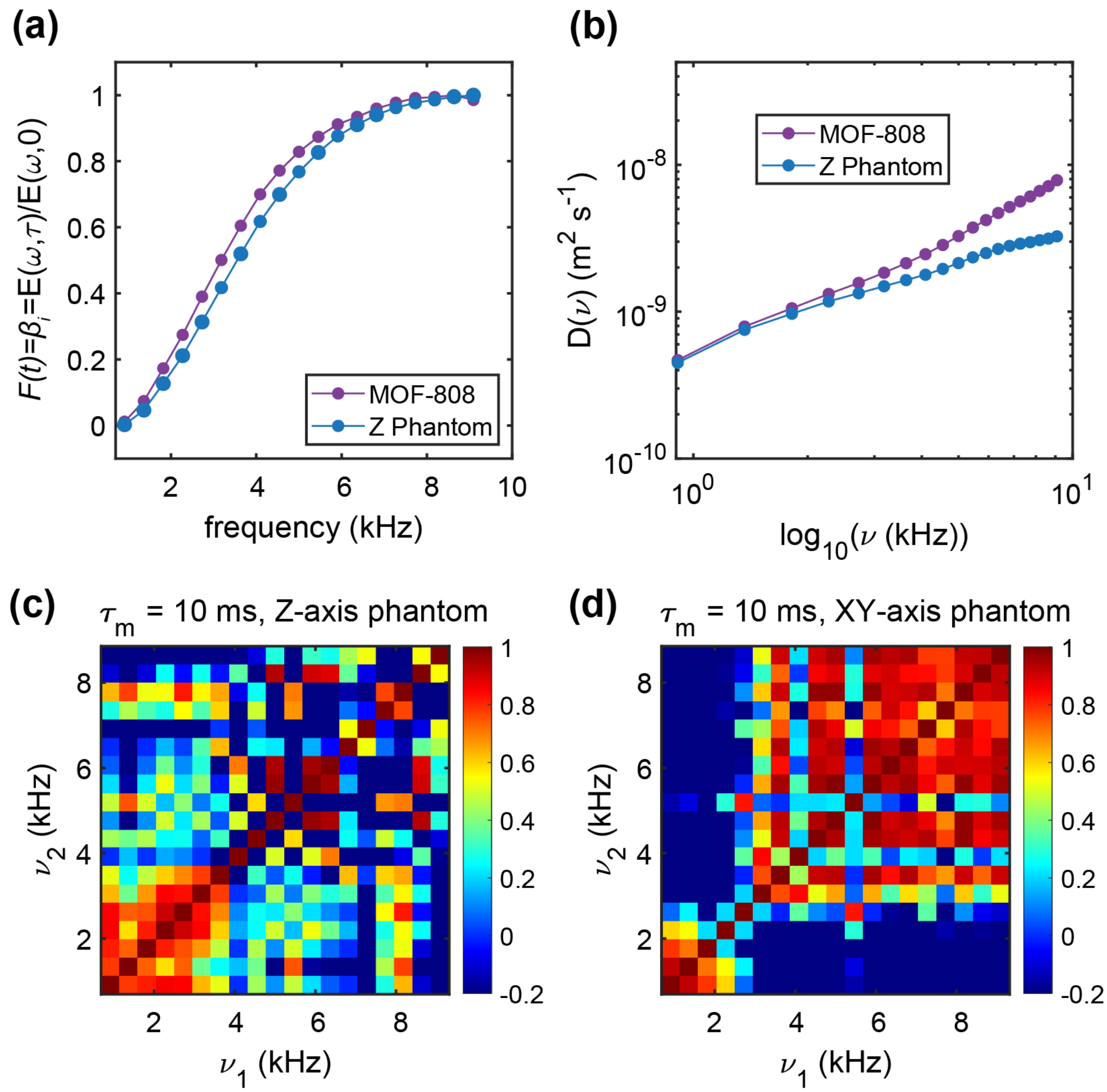

**FIG 3. Comparative OTOCs and diffusion spectra for one-dimensional control systems.** (a) OTOCs measured by MGSE for water confined in MOF-808 and for a control phantom of water confined in xylem bundles of celery, which form parallel cylindrical channels approximately 5 mm in diameter. (b) Corresponding diffusion power spectra, showing differences in frequency dependence between confined and one-dimensional geometries. (c) MGSE correlation plot for the cylindrical phantom oriented parallel to the magnetic field gradient (Z-axis), showing primarily Brownian motion. (d) MGSE correlation plot for the same phantom oriented in the XY plane, revealing confinement effects as increased high-frequency mixing.

Having established the experimental framework (Figs. 1–2) and validated its performance using a well-controlled test system (Fig. 3), we now apply the method to probe the correspondence between classical diffusion dynamics and quantum-mechanical OTOC principles. To this end, we use the MGSE-based OTOC measurement to reveal an entropy modulation effect that can be interpreted through a quantum-mechanical extension of Noether's theorem [11,13]. In this framework, the symmetry constraints that ensure energy conservation apply in analogous form to both physical diffusion and spin phase evolution; however, these conditions are often easier to satisfy—and to quantify—using macroscopic analogues, as in the present work.

Figure 4 demonstrates the entropy modulation effect that arises when the gradient–time wave vector used to measure physical-diffusion OTOCs is modulated in a nonlinear, periodic manner (see Methods section). The gradient modulation generates distinct spin coherences and, at each modulation frequency, selectively addresses specific proton spin sub-ensembles. The effect is negligible for free water, underscoring the role of the confining crystalline network. Entropy growth is quantified by the incremental change

$$\Delta S(\nu,t) = \ln D(\nu,t)/D(\nu,0) \tag{17}$$

as derived in the Supplementary Text [26,61–67]. Physically, high-frequency diffusion modes appear when water is confined to increasingly smaller physical domains, thus $D(\nu,t)$ is greater than $D(0,t)$ for confined systems. As a direct consequence, since entropy is related to heat flow through $\Delta Q = T\Delta S$, integration over frequency can yield the total $\Delta Q/T$ associated with the measured entropy change [61,66,68].

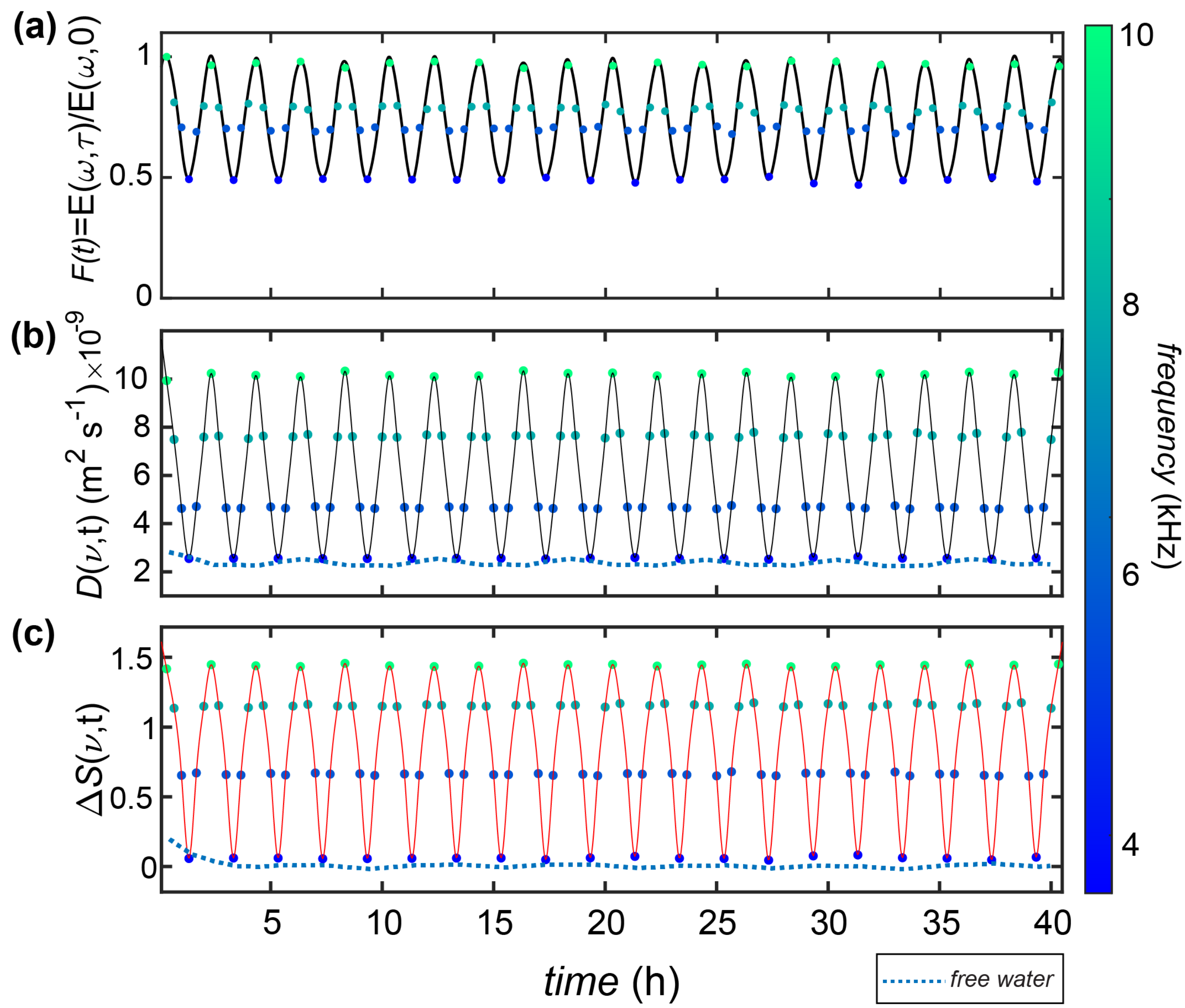

**FIG 4. Experimental demonstration of periodic entropy modulation, reflecting changes in accessible microstates through frequency-selected diffusive modes.** (a) OTOC as a function of frequency for water confined in MOF-808. (b) Corresponding diffusion spectrum obtained from (a). (c) Entropy change as a function of frequency, calculated from (b) using $D(\nu, t) \propto \ln S(\nu, t)$, which follows from the definition of incremental entropy change in Eq. (17) as derived in the Supplementary Text. This effect is not observed for free water, whose diffusivity and entropy remain near their infinite-time values (dashed blue line in (b)–(c)).

The frequency-resolved diffusion and entropy data in Fig. 4(b)–(c), derived from the OTOC in Fig. 4(a), highlight how the geometry of confinement and corresponding diffusive-mode selection govern the entropy modulation effect. Variation of the inter-echo dephasing time selects states of varying entropy in the water. During these times, diffusive disorder increases entropy to degrees set by confinement, which in turn determines the number of accessible microstates. Here, modulation frequencies span 3.5–9.5 kHz. The high-frequency (9.5 kHz, green) data markers correspond to the largest measured diffusivities $D(\nu, t)$ and associated entropy changes $\Delta S(\nu, t)$. This reflects the selection of the most spatially localized protons, which are those with the greatest number of accessible microstates. In contrast, the low-frequency (3.5 kHz, blue) markers correspond to the smallest $D(\nu, t)$ and $\Delta S(\nu, t)$, as these frequencies predominantly select water populations within the MOF whose motion is bulk-like, with minimal confinement and a correspondingly smaller state space; these values align with the featureless dashed line for free water. Physically, the observed frequency dependence follows from the fact that diffusion in a confined geometry is an eigenvalue problem, with additional modes emerging as confinement increases. In MOF-808, these eigenmodes map directly onto the pore topology, with localized modes corresponding to aqueous protons in small cages and delocalized modes spanning interconnected channels [69].

When assessing the useful work that can be extracted from an energy-conserving system, entropy must be considered, as it is the driver of dissipative losses [70,71]. An external "force" that restores a system to a high potential energy state does so by injecting low entropy, in the form of order or phase coherence that limits accessible microstates, after which the system evolves naturally toward higher entropy as new microstates become available. This is the basic principle of a heat engine; by analogy, spin-containing material networks can exhibit related entropy dynamics. Information can be encoded into such systems by preparing subsets in superposition states via π/2 rf pulses, followed by controlled evolution intervals. In this picture, entropy governs the diffusion processes that mediate energy dissipation. Boltzmann entropy ($S = k_B \ln \omega$) plays a role analogous to a classical force that drives dissipation, while partial but constrained knowledge of the system—arising from entanglement processes and couplings between superposition states—can modify this evolution. Information entropy is analogous to Boltzmann entropy, much as spin and physical diffusion can each carry analogous quantum information. Entropy thus reports the flow of information in quantum systems; notably, it scales with surface area, and its measurement fundamentally involves non-commuting operators [72,73].

The selection via π-pulse spacing between high and low frequencies thus alternates between high- and low-entropy modes, producing an ordered, periodic sequence of entropy growth and restoration—an effect reminiscent of discrete time-translation symmetry breaking in time crystals, though manifested here through diffusion modes in a confined molecular network [74–76]. In other contexts, this method offers a framework for studying non-equilibrium processes and macroscopic transport. Here, the observed oscillations correspond to mode selection within a system at thermal equilibrium; hence, they do not decay. However, repetition of this experiment in chaotic systems with changing architectures or concentration gradients would enable direct observation of how such processes alter available modes of diffusion and microstate evolution. In this scenario, the oscillation envelope may grow or decay at rates analogous to a macroscopic butterfly velocity, defined by the spacetime profile of $F(t)$ that marks the boundary of a commutator core. This velocity quantifies the speed at which initially localized perturbations influence distant degrees of freedom, complementing the Lyapunov exponent that describes the temporal growth rate. In chaotic systems with long-range interactions, sharply localized initial operators coupled with initially strong confinement boundaries may lead to hybrid behavior of $F(t)$, where exponential growth or decay is modulated by oscillatory effects.

## IV. DISCUSSION

In this work, we demonstrated that water molecules confined within the crystalline lattice of MOF-808 can emulate aspects of a quantum information processor. We developed and implemented an NMR-based approach for measuring the out-of-time-order correlator (OTOC) for physical diffusion and extended this framework to probe multidimensional correlations, enabling the observation of entanglement-like processes mediated by diffusive exchange. This approach provides a route to studying information scrambling and coherence dynamics in a macroscopically classical system.

By employing modulated gradient spin echo (MGSE) experiments, we accessed both position (via magnetic field gradients) and momentum (via the velocity autocorrelation function), thereby reconstructing the OTOC for proton motion in a spatially confined environment. We found that the emergence of frequency-selective excitation and entropy oscillations depends critically on structural confinement within MOF pores—effects absent in free water. These findings show that ensemble-averaged classical diffusion, when analyzed through MGSE and OTOCs, can display dynamics formally analogous to those of quantum circuits. More broadly, the confinement-controlled eigenmode spectrum suggests a hydrodynamic handle on scrambling rates. In structurally evolving systems, such as those with gradients or dynamic disorder, one expects mode-resolved OTOC to develop growth and decay envelopes governed by an effective butterfly velocity that bounds spatial perturbation spread, complementary to Lyapunov-like temporal growth.

Periodic rf driving, combined with gradient modulation, allowed observation of entropy dynamics: a sequence of entropy growth and reversal driven by frequency-selection of diffusion modes. This behavior parallels the operation of a heat engine, where external coherence inputs (rf pulses) inject low entropy into the system, and subsequent diffusion redistributes it. The entropy dynamics were quantified through frequency-resolved diffusion spectra, establishing a link between the thermodynamics of the spin system and classical heat exchange laws. In this setting, the relevant timescales are set by geometry-dependent hydrodynamic diffusion of aqueous protons, rather than by the $T_2$ coherence limits typical of strongly dipolar-coupled solids. Moreover, the topology and chemistry of the MOF pores provide a structural control layer over spin interactions, offering a hydrodynamic means to tune entropy modulation and OTOC dynamics via confinement geometry, molecular environment, and external driving fields.

In the presence of dipolar coupling, spin evolution acquires an additional transport term that is mathematically equivalent to a diffusion equation, termed 'spin diffusion' in the NMR literature to indicate dipole-mediated transport of magnetization. In the high-temperature, secular limit, the dipolar Hamiltonian reduces to a form mathematically equivalent to the diffusion (heat) equation for spin magnetization — this is the essence of the Bloembergen–Purcell–Pound/Redfield/Bloch-Torrey family of models, and the spin diffusion equation is written precisely this way in seminal texts by Abragam and Slichter [77,78]. In the language of classical transport, "diffusion" is any second-order Laplacian-governed spreading process (heat, particles, probability amplitudes, etc.); we have dealt with this classical form in the present work. However, we note that the multidimensional OTOC measurement used to track exchange in the macroscopic context may readily be applied to the study true entanglement processes in interacting spin chains that have been studied with other NMR implementations of OTOCs as "quantum simulator" systems [18].

## V. CONCLUSIONS

This work illustrates a convergence between classical statistical mechanics and quantum dynamics, showing that structure-driven classical dynamics can reproduce key operational features of quantum evolution. By employing MGSE-based NMR measurements of OTOCs, in one and two dimensions, we established that confined spin ensembles can exhibit entropy dynamics governed by periodic, frequency-selected diffusion modes. Frequency-selected diffusion reveals sustained, geometry-controlled entropy modulation absent in free water, confirming the confinement-based origin of the effect. Because the measurement fundamentally probes non-commuting operators, the entropy readout reflects information flow rather than ordinary relaxation. These findings define an experimental regime where quantum–classical analogies become quantitatively testable equivalences. In such systems, the interplay of coherence, entropy, and confinement provides both a conceptual link between disciplines and a practical framework for probing and manipulating information flow in physical materials. More broadly, these results establish porous frameworks and tabletop NMR as accessible platforms for exploring universal features of scrambling in condensed matter systems, bridging classical transport and quantum information dynamics.

**ACKNOWLEDGMENTS**

S.N.F. gratefully acknowledges support as a Pines Magnetic Resonance Center postdoctoral fellow and insightful conversations with N. Frank and B. King. S.K. would like to acknowledge funding from the U.S. Department of Energy (DOE) Office of Basic Energy Sciences Award No. DE-SC0019215. V.J.W. acknowledges salary support from 4R00GM140338-02 from the National Institute of General Medical Sciences at the National Institutes of Health.

**DATA AVAILABILITY**: The data that support the findings of this study are available from the corresponding author upon reasonable request.

**COMPETING INTERESTS**: The authors declare no competing interests.

Supplemental Material for

# Out-of-time-order correlators bridge classical transport and quantum dynamics

Sophia N. Fricke *et al.*

*Corresponding author. Email: snfricke@berkeley.edu

**This PDF file includes:**

Supplementary Text

Figures S1–S2

Supplementary References

**Supplementary Text**

***Entropy changes from diffusion-induced phase decoherence***

The modulated gradient spin echo (MGSE) signal measured in this work can be written as

$$E(\boldsymbol{q}, t) = \langle e^{i\phi(\boldsymbol{q},t)} \rangle \tag{S1}$$

where $\phi(t)$ is the accumulated phase resulting from spin motion in the applied gradient waveform $\boldsymbol{q}(t)$ [1,2]. For motion along the gradient direction, the phase may be expressed as

$$\phi(t) = \int_0^t \boldsymbol{q}(t')\, v(t')\, dt', \tag{S2}$$

with $v(t)$ the velocity component along the gradient direction. Angle brackets denote an ensemble average over spin trajectories. Here we adopt the Gaussian phase approximation, which is valid when phase accumulation arises from many uncorrelated or weakly correlated microscopic steps, such that $\phi(t)$ is well described as a zero-mean Gaussian random variable [3]. Under the Gaussian approximation, the signal attenuation is determined by the phase variance,

$$E(\boldsymbol{q}, t) = \exp\left[-\frac{1}{2}\,\mathrm{Var}(\phi)\right]. \tag{S3}$$

The phase variance may be written in the frequency domain as

$$\mathrm{Var}(\phi) = \int_0^\infty |\boldsymbol{q}(\nu)|^2 D(\nu, t) d\nu, \tag{S4}$$

where $\boldsymbol{q}(\nu)$ is the Fourier transform of the effective gradient modulation and $D(\nu, t)$ is the frequency-dependent diffusion spectrum (or velocity power spectrum) probed by the MGSE sequence. All sample- and time-dependent dynamical information enters through $D(\nu, t)$; the filter function $q(\nu)$ is fixed by the pulse sequence and hardware setup.

To quantify the spreading of phase information induced by diffusion, we associate an information entropy with the random phase variable $\phi$. For a continuous random variable with probability density $P(\phi)$, the Shannon (differential) entropy [4–7] is defined as

$$S_\phi = -\int P(\phi) \ln P(\phi)\, d\phi. \tag{S5}$$

For a Gaussian distribution with variance $\sigma_\phi^2 = \mathrm{Var}(\phi)$, this entropy is given analytically by

$$S_\phi = \frac{1}{2}\ln\left(2\pi e \sigma_\phi^2\right) = \frac{1}{2}\ln(2\pi e\,\mathrm{Var}(\phi)) \tag{S6}$$

where $e$ is Euler's number. Substituting Eq. (S4) yields

$$S_{\phi} = \frac{1}{2}\ln\left(2\pi e \int_{0}^{\infty} |q(\nu)|^{2} D(\nu,t)d\nu\right). \tag{S7}$$

Because differential entropy is defined only up to an additive constant and depends on units, we focus on entropy changes relative to a reference condition (e.g., an initial time $t = 0$) [8–10]. The entropy change is

$$\Delta S_{\phi}(t) = S_{\phi}(t) - S_{\phi}(0) = \frac{1}{2}\ln\left(\frac{\mathrm{Var}(\phi,t)}{\mathrm{Var}(\phi,0)}\right). \tag{S8}$$

Using the spectral representation,

$$\Delta S_{\phi}(t) = \frac{1}{2}\ln\left(\frac{\int_{0}^{\infty}|\boldsymbol{q}(\nu)|^{2}D(\nu,t)d\nu}{\int_{0}^{\infty}|\boldsymbol{q}(\nu)|^{2}D(\nu,0)d\nu}\right). \tag{S9}$$

For fixed experimental filtering $\boldsymbol{q}(\nu)$, entropy changes at a given frequency $\nu$ are directly tied to changes in the diffusion spectrum. Absorbing the factor of ½ into the entropy normalization (since only relative changes are considered), we obtain the frequency-resolved entropy change reported in the main text:

$$\Delta S(\nu,t) = \ln\left[\frac{D(\nu,t)}{D(\nu,0)}\right]. \tag{S10}$$

Thus, the relative entropy $\Delta S(\nu,t)$ quantifies the loss of phase information associated with diffusion-induced decoherence in a specified dynamical frequency band. It is an information theory form of entropy derived directly from experimentally measured quantities and does not rely on assumptions about thermodynamic equilibration or microscopic irreversibility.

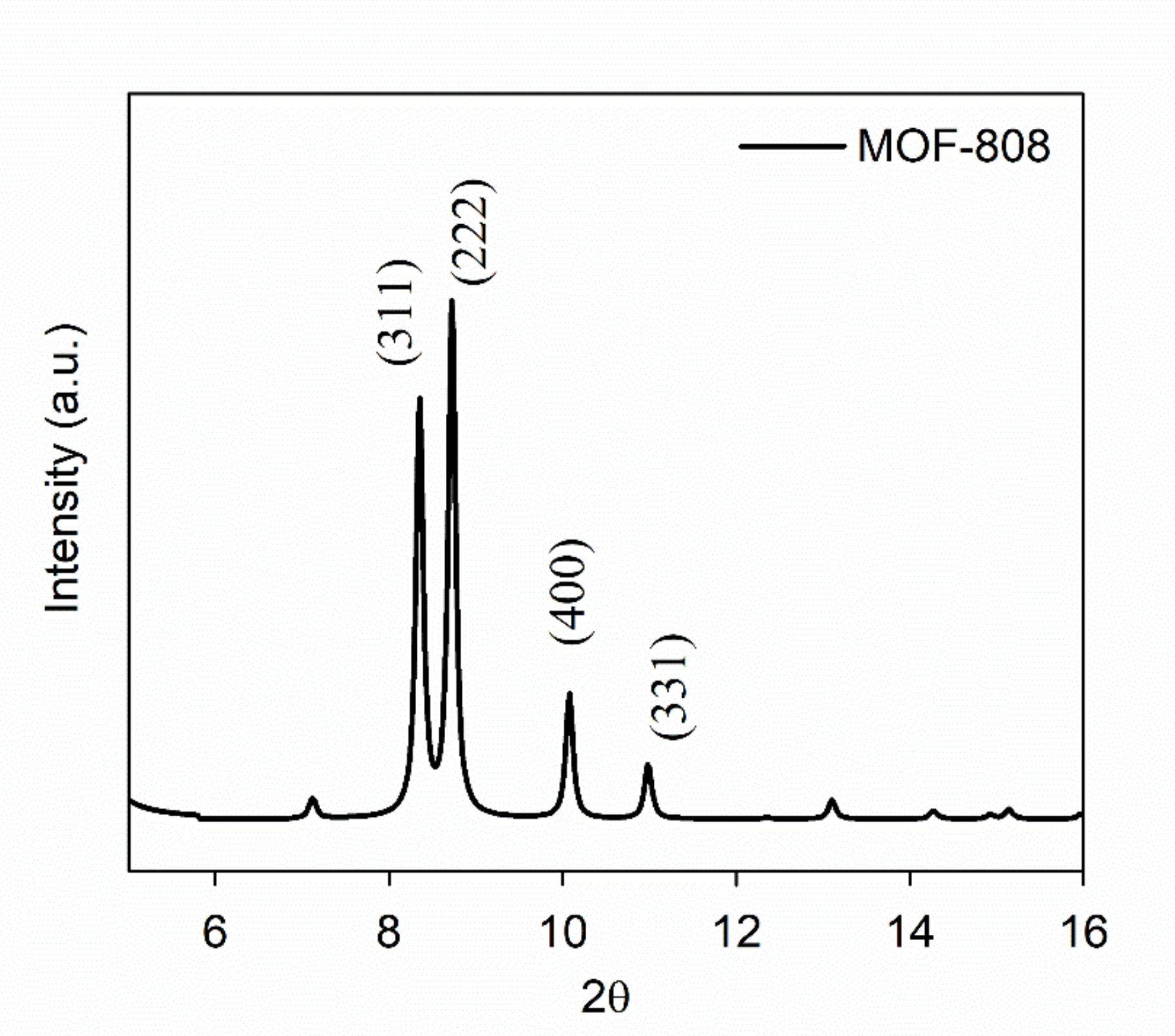

Figure S1. Powder X-ray diffraction pattern of MOF-808(Zr), recovered post-NMR measurement.

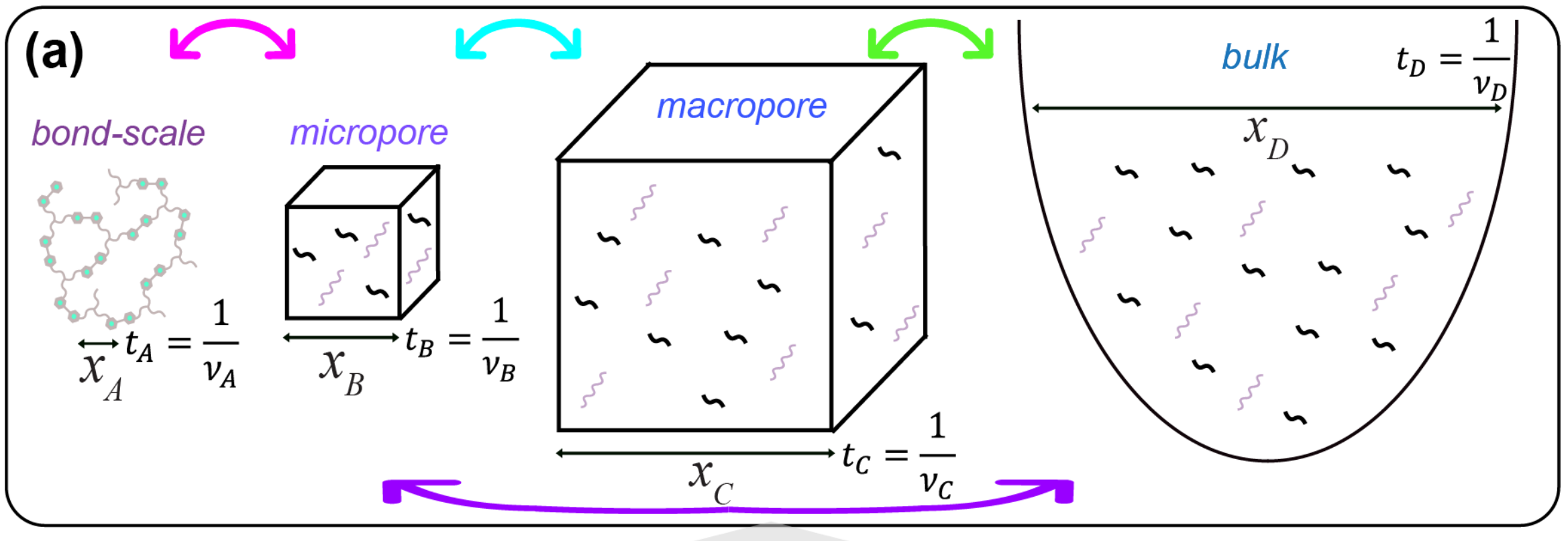

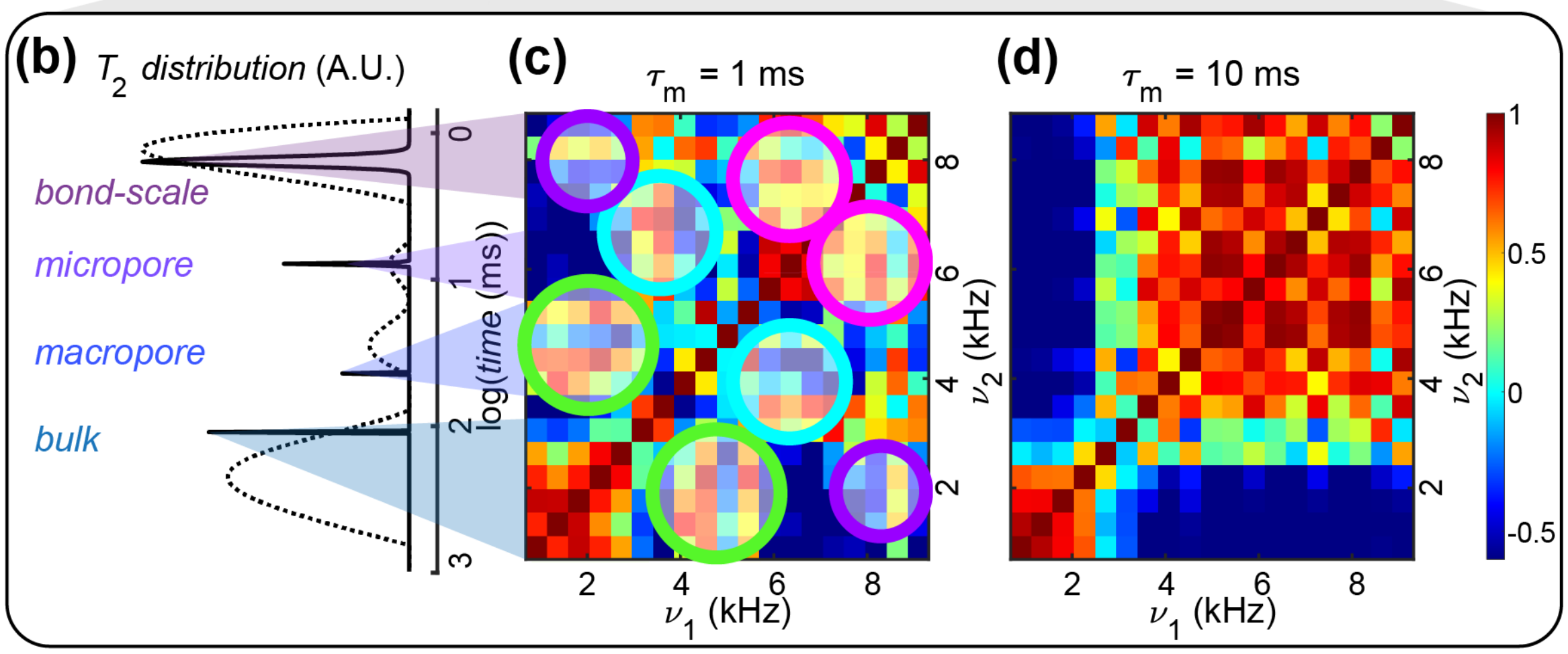

Figure S2. Multidimensional OTOCs report correlations via spatiotemporal exchange networks in water confined in MOF-808. (a) Schematic of water molecules experiencing motion in four distinct domains: bond-scale, micropore, macropore (intercrystallite), and bulk water. (b) $T_2 = \frac{1}{R_2}$ relaxation time distributions obtained by Laplace inversion (dashed) and matrix pencil method (solid). Peaks correspond to the four domains in (a), spanning over three orders of magnitude in relaxation times [11–14]. (c) 2D OTOC correlation plot for a 1 ms mixing time, showing discrete off-diagonal exchange between specific domains, e.g., between macropore and backbone-confined water (~2 and ~9 kHz). Colored circles indicate "jumps" corresponding to exchange between domains indicated by arrows of corresponding colors in (a). (d) 2D OTOC correlation plot for a 10 ms mixing time, showing widespread off-diagonal exchange, indicating extensive proton migration across domains.

Supplementary References